\documentclass{JINST}

\usepackage{float,verbatim}
\usepackage{subfigure} 
\usepackage{verbatim}            

\newcommand{\bit}{\begin{itemize}}
\newcommand{\eit}{\end{itemize}}

\title{Performance of the LHCb muon system with cosmic rays}
\author{
M.~Anelli$^a$, R.~Antunes Nobrega$^b$, G.~Auriemma$^b$$^c$, W.~Baldini$^d$, G.~Bencivenni$^a$ , R.~Berutti$^e$$^f$, V.~Bocci$^b$, N.~Bondar$^g$, W.~Bonivento$^e$\thanks{Corresponding author. EMail:Walter.Bonivento@cern.ch}, B.~Botchin$^g$, S.~Cadeddu$^e$, P.~Campana$^a$, G.~Carboni$^h$$^j$, A.~Cardini$^e$, M.~Carletti$^a$, P.~Ciambrone$^a$,  E.~Dane'$^a$,  S.~De~Capua$^h$$^u$,  C.~Deplano$^e$, P.~De Simone$^a$, F.~Dettori$^e$$^f$, A.~Falabella$^d$$^i$, F.~Ferreira Rodriguez$^k$, M.~Frosini$^l$$^m$, S.~Furcas$^a$$^n$,  G.~Graziani$^l$ , L.~Gruber$^o$$^p$, A.~Kashchuk$^g$, A.~Lai$^e$, G.~Lanfranchi$^a$, M.~Lenzi$^l$, O.Levitskaya$^g$, K.~Mair$^o$, O.~Maev$^g$, G.~Manca$^e$, G.~Martellotti$^b$, A.~Massafferri Rodrigues$^j$$^q$, R.~Messi$^h$$^j$, F.~Murtas $^a$, P.~Neustroev$^g$, R.G.C.~Oldeman$^e$$^f$, M.~Palutan$^a$, G.~Passaleva$^l$, G.~Penso$^b$$^r$, A.~Petrella$^d$$^i$, D.~Pinci$^b$, S.~Pozzi$^d$$^l$, G.~Sabatino$^h$$^j$, B.~Saitta$^e$$^f$, R.~Santacesaria$^b$,  E.~Santovetti$^h$$^j$, A.~Saputi$^a$, A.~Sarti$^a$, C.~Satriano$^b$$^c$, A.~Satta$^h$, M.~Savri\'e$^d$$^i$, B.~Schmidt$^o$,  T.~Schneider$^o$, A.~Sciubba$^a$$^r$, P.~Shatalov$^s$,  S.~Vecchi$^d$, M.~Veltri$^l$$^t$, S.~Volkov$^g$, A.~Vorobyev$^g$ \\
\llap{$^a$}Laboratori Nazionali di Frascati dell'INFN, Frascati, Italy \\
\llap{$^b$}Sezione INFN, Roma, Italy \\
\llap{$^c$}University of Basilicata, Potenza \\
\llap{$^d$}Sezione INFN, Ferrara, Italy \\
\llap{$^e$}Sezione INFN, Cagliari, Italy \\
\llap{$^f$}Universit\`{a} di Cagliari, Cagliari, Italy \\
\llap{$^g$}Petersburg Nuclear Physics Institute, Gatchina, St-Petersburg, Russia \\
\llap{$^h$}Sezione INFN, Roma Tor Vergata, Italy  \\
\llap{$^i$}Universit\`{a} di Ferrara, Ferrara, Italy \\
\llap{$^j$}Universit\`a di Roma Tor Vergata\\
\llap{$^k$}Instituto de F\'\i sica - Universidade Federal do Rio de Janeiro (IF-UFRJ), Rio de Janeiro, Brazil \\ 
\llap{$^l$}Sezione INFN, Firenze, Italy \\
\llap{$^m$}Universit\`{a} di Firenze, Firenze, Italy \\
\llap{$^n$}now at Sezione INFN, Milano, Italy \\
\llap{$^o$}European Organisation for Nuclear Research (CERN), Geneva, Switzerland \\
\llap{$^p$}Technische Universitat Wien, Austria \\
\llap{$^q$}Centro Brasileiro de Pesquisas F\'\i sicas (CBPF), Rio de Janeiro, Brazil\\
\llap{$^r$}Universit\`{a} di Roma "La Sapienza"\\
\llap{$^s$}ITEP Moscow \\
\llap{$^t$}Universit\`{a} di Urbino, Urbino, Italy \\
\llap{$^u$}now at EPFL, Lausanne, Switzerland \\
}

\abstract{The LHCb Muon system performance is presented using  cosmic ray events collected in 2009. These events allowed to test and optimize the detector configuration  before the LHC start.  The space and time alignment and the measurement of chamber efficiency, time resolution and cluster size are described in detail. The results are in agreement with the  expected detector performance. }


\begin{document}
\section{Introduction}
\label{sec:intro}

The main purpose  of the  LHCb muon detector~\cite{bib:lhcbpaper}   is to provide the  LHCb experiment with a  trigger for b-hadron decay channels containing muons in the final state. Moreover, it is the main sub-detector providing off-line  muon identification. It consists of  five stations, M1 to M5, equipped with multi-wire proportional chambers (MWPC), with the exception of the inner part of the first station equipped with triple-GEM detectors.    For  triggering, the detector  has to be highly efficient, more than 95~\%, on muons within a time window smaller than 25ns to unambiguously identify the LHC bunch crossing.  The detector and its associated readout electronics were  optimized for this goal  and test beams measurements  with prototype detectors confirmed the expected performance.\\
However, the construction of a very large system (1380 chambers with 122,000 readout channels), assembled in  different productions sites during several years, and with some technical details different from site to site, is such that some  chamber to chamber  non-uniformity  can be  expected. In addition, the operation of this very large system can  be affected by problems  not present when testing one chamber at the time in the lab or with test beams.
Therefore, it is crucial to  assess the system performance as a whole  in order to confirm the expected results.\\
 At the beginning of 2009 a large  data sample of cosmic rays was collected. Data were taken in different detector conditions and a recursive  optimization process lead to a long  data taking period whose results are described in this paper. \\
The main optimization issues concerned  the space and the time alignment. The forward geometry of the LHCb experiment is not optimal to detect cosmic rays. In particular the inner regions of the muon detector would require high statistics of almost horizontal tracks. On the other hand, cosmic rays permit a good calibration in the outer regions where muons from LHC interactions are scarce.\\
In order to assess the  system performance, the measurement of chamber efficiency, time resolution and cluster size  are also described here. As the main goal of this work was to assess the performance of the detector chambers, methods have been devised to separate contributions to the  efficiencies and resolutions measured with cosmic rays that are linked to the geometry of the system from contributions coming  from the detector performance itself. 

\section{The LHCb muon system}
\label{sec:muondet}
\begin{figure}[ht]
\centering{
\includegraphics[width=0.5\textwidth]{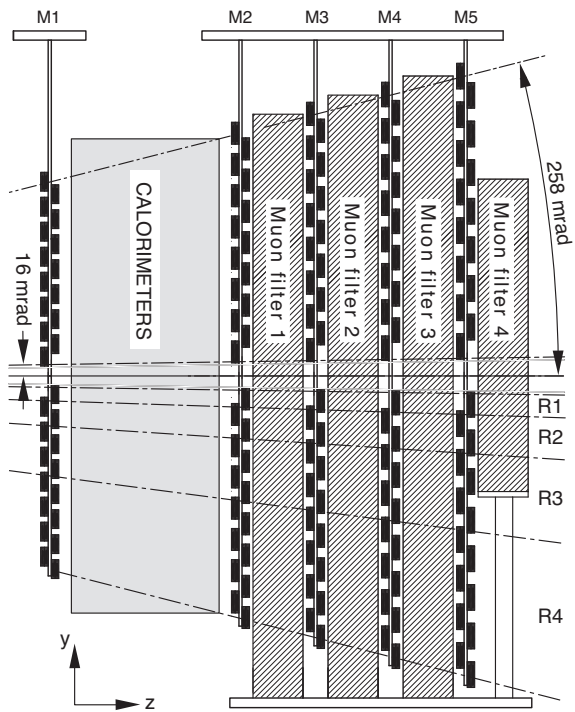}
\caption{
Side view of the LHCb muon system. The LHCb reference system is a right-handed coordinate system with the nominal collision point at the origin, with the  $z$ axis  defined by the beam axis and directed from the VELO to the muon system.}
\label{fig:sideview}			 
 }
\end{figure}

\begin{figure}[ht]
\centering{
\includegraphics[width=1.\textwidth]{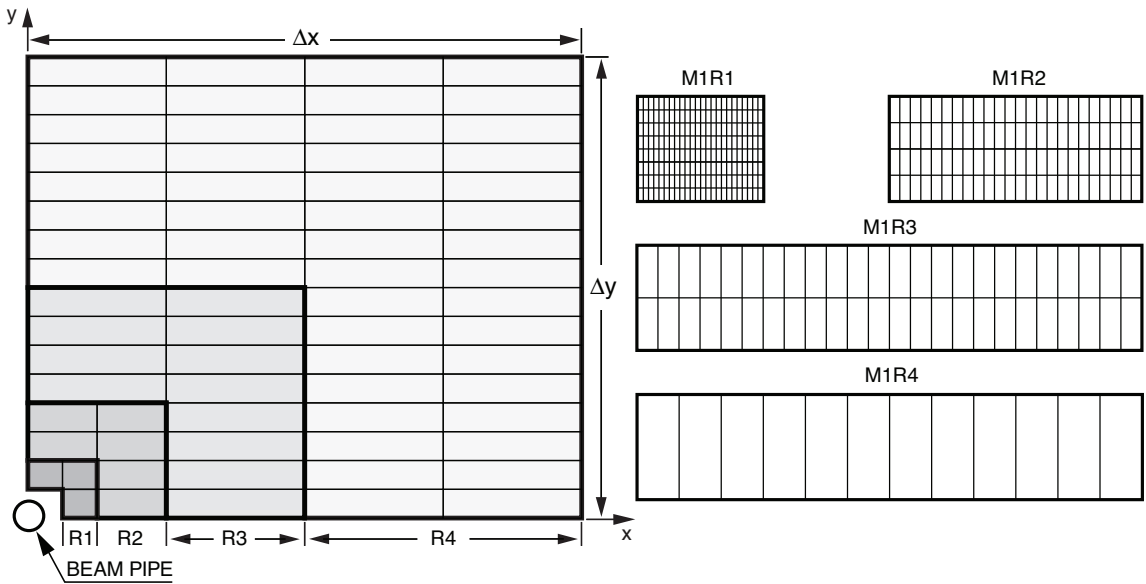}
\caption{Left: front view of a quadrant of a muon station. Each rectangle
represents one chamber. Each station
contains 276 chambers. 
Right: division into logical pads of four chambers belonging to the
four regions of station M1. In each region of stations M2-M3
(M4-M5)  the number of pad columns per chamber is double (half) the number
in the corresponding region of station M1, while the number of pad rows
per chamber is the same.}
\label{fig:radius}			 
}
\end{figure}

\begin{figure}[hbt]
\begin{center}
\includegraphics[width=0.49\textwidth]{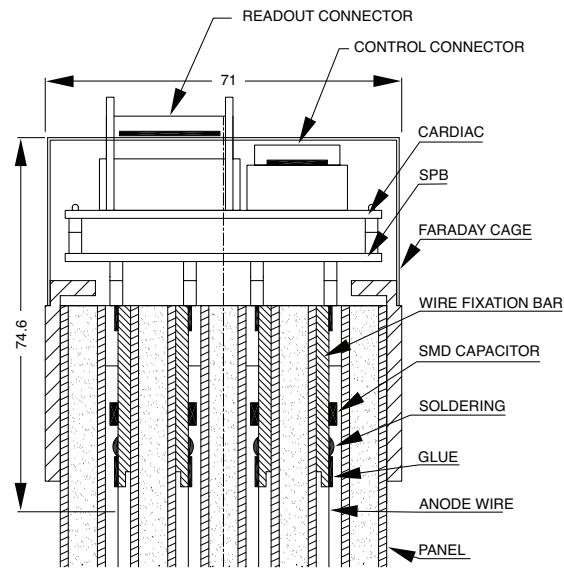}
\caption{Cross section of a wire chamber showing the four gas gaps
and the  connection to the readout electronics. SPB: Spark
Protection Board; {\sc cardiac}: FE Electronics Board. In this case 
the hardwired OR 
forming the two double gaps (see text) is achieved in the SPB.}
\protect\label{fig:xsection}
\end{center}
\end{figure}

\begin{figure}[htb]
\begin{center}
\includegraphics[width=0.49\textwidth]{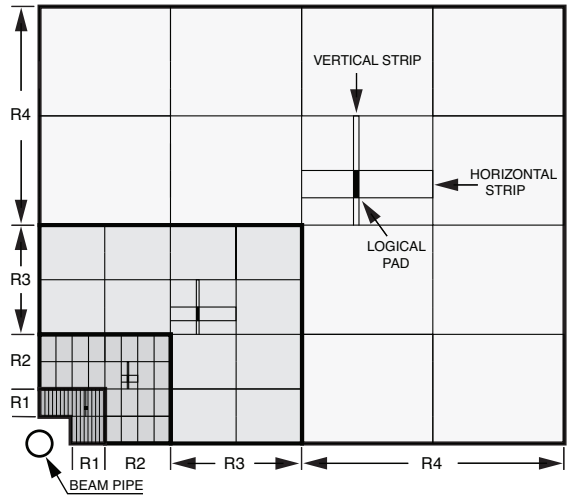}
\caption{ 
Front view of one quadrant of stations M2 and M3 showing the
partitioning into sectors. 
In one sector of each region a horizontal and a 
vertical strip are shown. The intersection of a horizontal and a vertical
strip defines a logical pad (see text). 
A Sector of region R1 (R2, R3, R4) contains 
8 (4, 4, 4) horizontal strips and 6 (12, 24, 24) vertical strips.}
\protect\label{fig:sectors}
\end{center}
\end{figure}

LHCb~\cite{bib:lhcbpaper}  is an experiment dedicated to heavy flavour physics at the LHC. Its primary goal is  to look for  indirect evidence of new physics in CP violation and rare decays of beauty and charm hadrons. 
The LHCb apparatus is a single-arm 
forward spectrometer, consisting of  a series of sub-detector systems, aligned as in a fixed-target geometry, along the beam axis. 
It includes a silicon-strip Vertex Locator (VELO) centered on the interaction point for precise vertex reconstruction. A dipole warm magnet provides the bending for momentum measurement. Tracking is insured by tracking stations using silicon strips (TT,IT) and straw tubes (OT) before and after the dipole. Particle identification is provided by two RICH detectors, by an electromagnetic and hadron calorimeter system (ECAL and HCAL) and by the Muon Detector. \\
The muon system  is composed of five stations (M1-M5) of rectangular shape, placed along the beam axis. 
As shown in figure~\ref{fig:sideview}, stations M2 to M5 are placed downstream the calorimeters and are interleaved with iron absorbers 80 cm thick to select penetrating muons.  Station M1 is instead located in front of the calorimeters and is
used to improve the transverse momentum  measurement in the first level hardware trigger L0.  \\
The geometry of the muon detector was designed in order to fulfill requirements of both  performance and  easy access to the detector itself.
On each station, 276  chambers are mounted on aluminium supporting walls at four different distances  ($\pm$58.5~mm and $\pm$142.5~mm from the station middle plane)  in order to provide with their sensitive area a hermetic geometric acceptance  to high momentum particles coming from the interaction point. 
In addition, the chambers of different stations are placed so that they form projective towers pointing the interaction point. 
Each  station consists of two mechanically independent halves (called A and C side), hanging from a common rail,
that can be opened to access the beam pipe and the detector chambers for maintenance. \\
       The detectors provide space point measurements of the tracks, providing binary (yes/no) information to the trigger processor  and to the data acquisition (DAQ). The information is obtained by
partitioning the detector into rectangular logical pads whose dimensions define the $x$, $y$ resolution. \\
The muon trigger is based on stand-alone muon track reconstruction and transverse momentum ($p_{T}$) measurement, with a 20\% resolution, 
and requires aligned hits in all five stations. Since the spectrometer dipole provides bending in the horizontal plane, the pad segmentation of muon chambers is finer in the horizontal direction than in vertical one, to allow a precise estimate of the momentum.  Stations M1 to M3 are used to define the track direction and to calculate the
$p_{T}$ of the candidate muon and therefore have a higher spatial resolution along
the $x$ coordinate (bending plane) than stations M4 and M5, whose  main purpose is the identification of penetrating particles. The dimensions of the M1 pads in the inner region of M1 station are 1~cm in $x$ and 2.5~cm in $y$. The pad vertical size is the same (apart from the projective increase) in all the other stations, while the $x$ size is two times smaller in station M2 and M3 and two times larger in M4 and M5. \\
The positioning of the chambers in the $x$-$y$
plane within a station is done in such a way as to
preserve as much as possible the full projectivity of
the logical layout. This is mandatory for a correct
execution of the L0-muon trigger algorithm and to
minimise the geometrical cluster size and geometrical
inefficiencies at the boundary of the chambers.
The logical layout is defined at the central
plane of the station and the sensitive area of each
chamber is sized as if it were at this plane. The
$x$ and $y$ positions of the centres of each chamber
within a station are obtained simply by positioning
each chamber centre so that it projects from the
interaction point  to its position in the logical
layout at the central plane of the station. In doing
so, the chambers in front of the supporting wall will overlap in
$x$ with their neighbours. The overlap however is
always less than half of one logical channel. Similarly,
the holes introduced between the chambers located behind the supporting wall are small, and are further limited
by the thickness of the chambers in $z$.
Viewed from the interaction point the total loss in
angular acceptance is less than 0.1\%. 
The corresponding $y$ overlaps are negligible due to the small
$y$ dimensions of the chambers. \\
     Each muon station is divided into
four regions, R1 to R4 with increasing distance from the beam axis. The linear dimensions of the
regions R1, R2, R3, R4, and their segmentation scale in the ratio 1:2:4:8, as shown in figure~\ref{fig:radius}. With this geometry,
the particle flux and channel occupancy are expected to be roughly of the  same order of magnitude over the four regions
of a given station. \\
      The trigger algorithm requires a five-fold coincidence between all the stations, therefore the
efficiency of each station must be $\ge$~99\% to obtain a trigger efficiency of at least 95\%, within
a time window smaller than 25 ns in order to unambiguously identify the bunch crossing (BX). \\
The necessary time resolution is ensured by a fast gas mixture, \(\mathrm{Ar/CO_{2}/CF_{4}}\) 40/55/5, and an optimized charge-collection
geometry both for the MWPC and the GEM detectors. Moreover, the chambers are composed of
four or two OR-ed gas gaps depending on station. In stations M2 to M5 the MWPC's are composed of
four gas gaps arranged in two sensitive layers with independent readout, as shown in figure~\ref{fig:xsection}. In station M1, R2 to R4  the MWPC have only two gas gaps to minimize the material in front of the electromagnetic calorimeter. In
region M1R1 two superimposed GEM chambers connected in OR are used. \\
Since spatial resolution and rate capability  vary strongly over the detectors, different readout techniques  are employed for the MWPC in different stations and regions.
      All the chambers are segmented into physical pads: anode wire pads, where the pads are formed by adding the analog signals coming from  a certain number of adjacent wires, or cathode pads, with a segmented cathode printed circuit board,  in the
MWPCs and anode pads, again with a segmented  printed circuit board, in the GEM chambers. \\
Each physical pad is read out by one front-end
(FE) electronics channel.
      The electronics includes flexible logical units performing the OR of a variable number of FE
channels following the requirements of the readout. 
Up to four adjacent physical pads are OR-ed by the FE electronics to
build a logical pad.  In the M1 station, where the foreseen channel occupancy is high, the signals from the logical
pads are sent directly to the trigger and DAQ. In most of the other regions, M2/3 R3/4 and M4/5 R2/3/4, several contiguous logical pads are further OR-ed to build
larger logical channels in the form of vertical and horizontal strips. The logical pads are then
reconstructed by the coincidence of two crossing strips, as shown in figure\ref{fig:sectors}. 
 However, in the high granularity regions R1-R2 of stations M2-M3 
a mixed readout was adopted: a narrow wire-strip defining the $x$ resolution  and a larger cathode pad
 defining the $y$ resolution are  the logical channels sent to the trigger and DAQ.
Logical pads are then obtained as an AND between wire and cathode pads.
     
\begin{figure}[ht]
      \centering
      \includegraphics[width=0.7\textwidth]{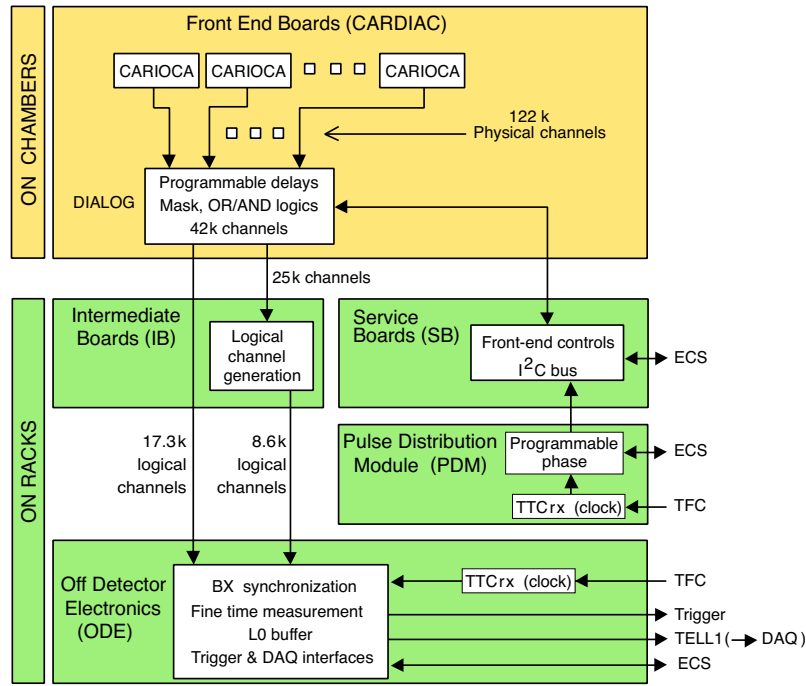}
\caption{Simplified scheme of the Muon electronics architecture.}
      \label{fig:elecscheme}
\end{figure}
Figure~\ref{fig:elecscheme} shows schematically the architecture of the Muon readout electronics. The task of the
electronics is twofold: to prepare the information needed by the Level-0 muon trigger and to send the
data to the DAQ system. The front-end (FE) CARDIAC boards house two eight channel ASIC's, each one  containing  a high bandwidth current amplifier, a shaper and a single threshold fast discriminator in leading edge mode,  processing the  122~k physical signals from the  chambers  (CARIOCA~\cite{bib:carioca}) and generate the 26~k logical-channel signals
 by suitable logical OR's of the physical
       channels (DIALOG~\cite{bib:dialog1}). This last step is in fact fully performed on the FE boards only in part of the detector and it is ended on special Intermediate Boards (IB) in regions where the logical channel spans more than one FE board. Eventually, the Off Detector Electronics (ODE) boards receive the signals from the logical channels. They are tagged with the number of the bunch crossing (BXID) and routed to the
       trigger processors via optical links without zero suppression. The fine time information inside the 25~ns gate, measured by a 4-bit TDC ASIC (SYNC~\cite{bib:dialog}) on the ODE boards, is added and the data
are transmitted via optical links to the TELL1 board~\cite{bib:tell1} and from the TELL1 to the DAQ system.

\section{Detector settings}
\label{sec:detreq}

\begin{figure}[ht]
  \centering
	\includegraphics[width=0.76\textwidth]{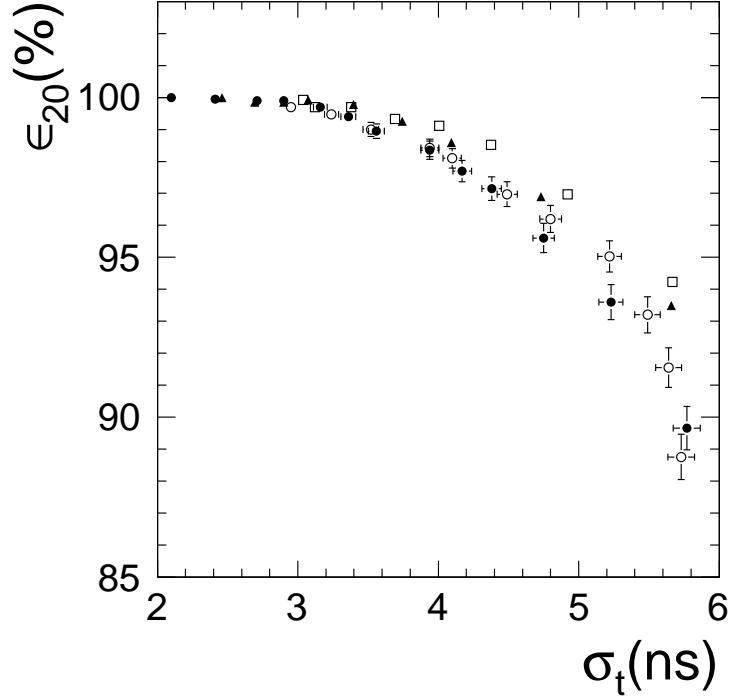}
	\caption{
 Efficiency  in 20~ns vs. time resolution (c) from a simulation of four-gap chambers (solid circles) and two-gap chambers (open circles) and test beam data for a M3R3 (pad readout) chamber (solid triangles) and a M5R4 (wire readout) chamber (open squares). }
	\label{fig:EffRMSvsThr}
\end{figure}

As described in section~\ref{sec:muondet}, the trigger requires the coincidence within the 25~ns LHC gate of signals from the five stations. 
 As a safety margin, the requirement of 95~\% efficiency (99~\% per station) was specified in a 20~ns window. 
Moreover, for an optimal trigger performance, the cluster size, i.e. the average number of pads yielding a signal per track in a given chamber, should not exceed 1.3-1.4 in stations M1 to M3. \\
Despite the different type of readout (wire or pad), from the point of view of performance the main difference is  the value of the readout pad capacitance to ground, which ranges from 50~pF to 245~pF, and affects the front-end amplifier sensitivity. The gas gaps and the wire pitch and radius are everywhere the same (except for the triple-GEM detectors) and therefore the induced charge  on the electrodes is also everywhere the same, at given high voltage.  
In terms of  detector performance, the muon detector has a crucial dependence on the  gas gain to threshold  ratio, since the  higher this ratio the better the time resolution (and efficiency), due to the reduction of the time walk effect. Unfortunately, a higher ratio  also implies   higher  cluster size,  higher collected charge at the anodes and  related aging  and  detector instability, such as the discharge probability, leading in the long run to detector failures. Therefore the performance optimization is a very delicate and careful work which has to take into account all these parameters. \\
The 20~ns detector efficiency is the main parameter qualifying chamber response in the LHCb muon system. However, this could not directly be  measured  with cosmic rays,  due to a non linear behavior of the TDC,  which will be discussed in section~\ref{subsubsec:timeAligCosmic}.  Indeed, since the  cosmic rays  are not  in time with the LHC clock, to measure the 20~ns detector efficiency 
one would have had to precisely measure the fine time all along the LHC gate, including the borders, for all hits associated to the track. Unfortunately the above mentioned TDC feature prevented us from  precise measurement right at  the LHC gate borders, compromising this measurement. \\
However, the 20~ns  efficiency  could  be determined in a indirect way by measuring the total efficiency in an infinite time window, the chamber time resolution and using results from laboratory tests~\cite{bib:test_beam} to make the connection between time resolution and  20~ns detector efficiency.  For comparison a chamber simulation was also performed using the drift chamber simulation program GARFIELD \cite{veenhof} for the gas mixture in use and a high voltage of 2.65~kV, corresponding to a gas gain of $10^{5}$. 
Figure~\ref{fig:EffRMSvsThr}  the 20~ns efficiency   vs. time resolution  obtained by simulation of four-gap chambers and two-gap chambers as well as test beam data for a M3R3 (pad readout) chamber  and a M5R4 (wire readout) chamber. \\
To preserve long term operation of the system, the MWPC's of the LHCb muon system should be operated at the lowest possible threshold compatible with electronic noise. For the cosmic ray data taking, the thresholds were set higher than foreseen for the  LHCb run at nominal luminosity in order to keep the noise level below 100~Hz per channel. The relatively large range of detector capacitance, together with the corresponding slewing effect in the front-end amplifier, is such that the same noise level is reached at quite different values of threshold if expressed in charge units. 
The set thresholds  ranged from  2.8~fC to about 11~fC  depending on stations and regions of the detector. \\
To equalize the gas gain to threshold ratio and therefore have the same efficiency everywhere no matter what the detector capacitance is, different high voltage values in each individual region and station~\cite{bib:anatoli} would have to be set.  However, for the cosmic ray acquisition runs and  the first very low luminosity LHC run, it was decided to start with the same value of the high voltage, 2.65~kV, for the whole detector, set to the highest 
calculated value among the different regions and stations compatible with the efficiency requirements and with  detector stability. Therefore, this high voltage settings leads in some stations and regions  to a larger than needed ratio of high voltage to threshold, i.e. to an expected 20~ns efficiency  beyond 99\%.  A more careful tuning of this ratio will be performed in the future.
\begin{figure}[ht]
      \centering
      \includegraphics[width=0.7\textwidth]{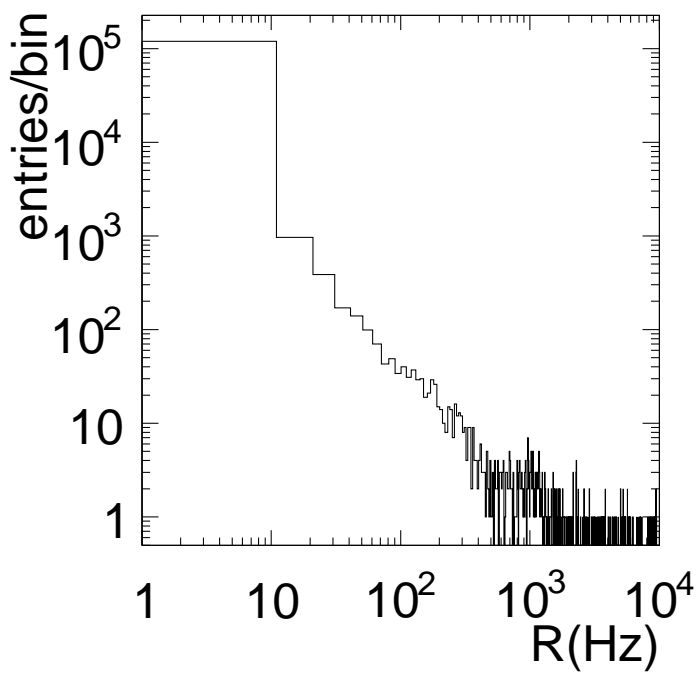}
\caption{Measured noise rate R(Hz)  for all  channels of the muon system.}
      \label{fig:NoiseMonitor}
\end{figure}
 Figure \ref{fig:NoiseMonitor} shows the noise rate for the whole muon system after threshold setting and shows 
that for \(99.3\;\%\) of all the channels the noise is below 100~Hz and that for \(99.8\;\%\) of the channels the noise rate 
is below \(\mathrm{1\;kHz}\).

\section{Data sample and track reconstruction}
\label{sec:trackrec}
\begin{figure}[ht]
  \centering
	\subfigure[]{
	\includegraphics[width=0.7\textwidth]{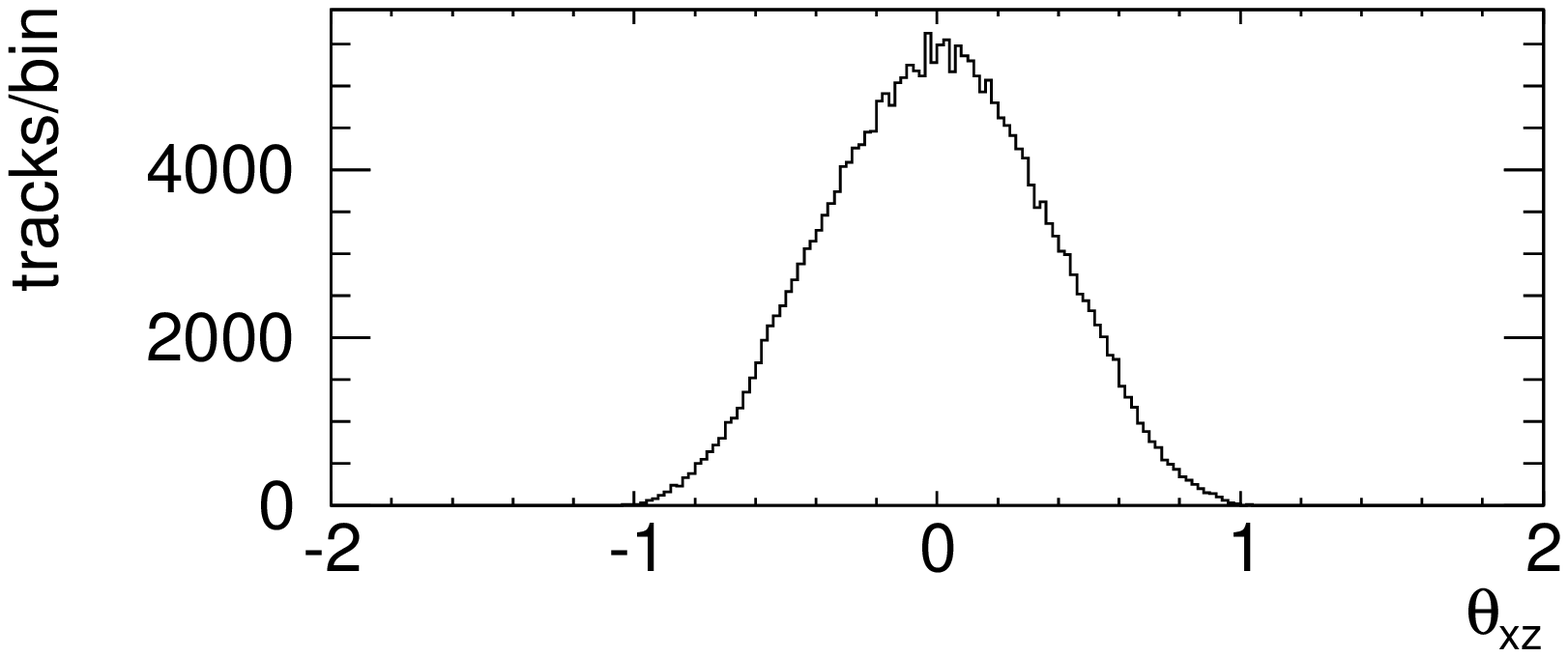}
	} \\
	\subfigure[]{
	\includegraphics[width=0.7\textwidth]{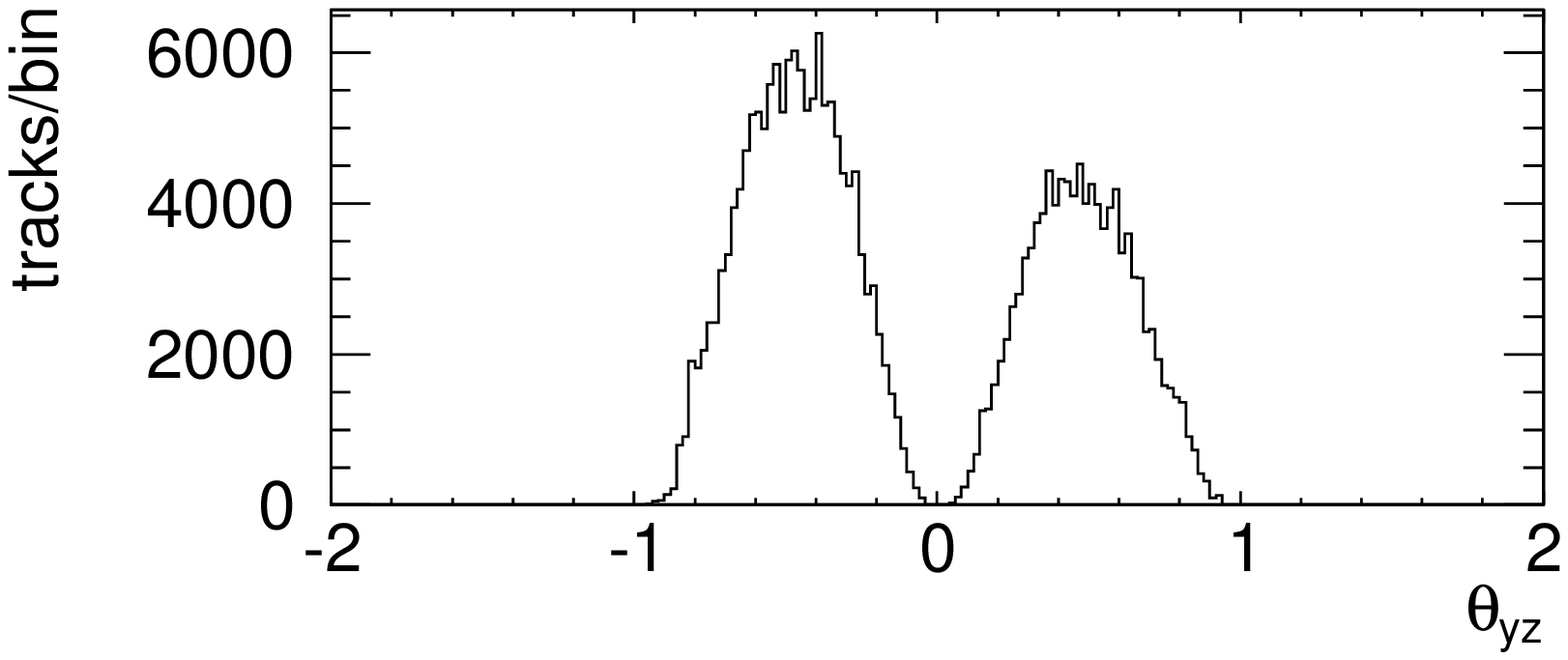}
	}
\caption{Angles (rad) in the horizontal plane $\theta_{xz}$ (a) and in the vertical plane $\theta_{yz}$ (b)  in the LHCb reference frame for reconstructed cosmic ray tracks.}
\label{fig::slopes}			 
\end{figure}

 The data sample for the analysis described in this paper consisted of 2.5~million  cosmic ray events  triggered by the   calorimeter, with the threshold set to detect  minimum ionizing particles. Events were acquired, according to a prescription of the DAQ team which affected all LHCb sub-detectors, in a time window of five LHC gates, i.e. 125~ns (\emph{ wide gate} in the following). This wide gate  indeed proved to be very useful since it allowed track reconstruction and a quite detailed  study of the time properties of signals, including cross-talk,  even before a precise time alignment of the detector took place. \\
Even if the LHCb geometry is not optimal for cosmic ray detection, the calorimeter trigger provided events at the rate of few Hz. \\
The  LHCb standard pattern recognition and tracking rely heavily on the position of the primary interaction vertex, and this makes it unsuitable to reconstruct cosmic ray tracks. Therefore two ad-hoc stand-alone  pattern recognition methods were developed. For  space and time alignment, time resolution and efficiency studies,  a neural network approach~\cite{bib:gipNN} was used, which is highly efficient and also allowed the reconstruction of multiple tracks per  trigger. Pattern recognition started from clusters of adjacent pads and the hit position in a station was determined from the cluster  barycenter. For  cluster size   and for a second analysis of the total efficiency, a  standard pattern recognition algorithm was used, looking for  the combination of more than three aligned hits (one per station) providing the best  fit, taking into account multiple scattering effects in the iron wall and in the calorimeters. In both cases the fit tracks were straight lines, given that no magnetic field was present in the muon chambers. \\
With the neural network approach, about 250,000 tracks were reconstructed with at least four hit stations.
Figure~\ref{fig::slopes} shows the track angles $\theta_{xz}$,  in the $xz$ plane, (a) and $\theta_{yz}$,  in the $yz$ plane, (b), in the LHCb reference frame; the two peaks at negative and  positive $\theta_{yz}$ in figure~\ref{fig::slopes}(b) correspond to  cosmic rays  going 
forward and backward in the apparatus. \\
As seen from figure~\ref{fig::slopes}, cosmic rays have a very different spatial and angular distributions compared  to  particles coming from the interaction point. The muon detector
is indeed built to be  hermetic for
tracks in acceptance coming from the interaction
point  by a suitable layout of the chambers in the stations. For this reason, 
cosmic rays  can go through un-instrumented 
areas of the detector, yielding a strong angular dependence of the efficiency, which in turn leads to potential systematic effects in the total efficiency determination and in the space alignment procedure, which should be carefully studied. 

\section{Space and time alignment}

Two essential ingredients to optimize the system performance are the space and time alignment, which are described hereafter and have to be performed before the efficiency and the resolution measurements.

\subsection{Space alignment}

The accurate spatial alignment of the muon detector is important to guarantee the design performance of trigger and  off line muon identification. 
Given the spatial resolution of the detector readout elements,
the needed alignment accuracy is driven by the trigger requirements in the inner regions of stations M1, M2 and M3. A precision of $\sim$1~mm in $x$ and $y$ directions is sufficient to guarantee the design specifications. The alignment requirements along $z$ are much less demanding due to the forward geometry of the experiment.
\subsubsection{Mechanical alignment}
During the installation the muon chambers were mounted on the supporting walls with a precision of 
$\sim$1~mm 
 centered on their nominal positions, calculated with respect to reference targets placed on top of each half station. 
The measured rotations were zero within the precision of 1~mrad.  
After  chamber installation, the muon filters and the half stations were closed around the beam pipe. Since the muon filters could not be completely closed because of mechanical tolerances of the iron blocks,  also the detector half stations were kept slightly open to avoid possible radiation damage. 
The opening of  each half is $\pm$5~mm at M1 and increases with the $z$ coordinate to preserve the projectivity of the muon chambers. 
The positions of the half stations with respect to the LHCb cavern reference were precisely surveyed using  four reference targets on each side, and the values were stored in the geometry database that is used by the reconstruction program to define the absolute hit coordinates. The average $x$ coordinate of the inner edges of the ten half stations, as measured by the survey, as a function of the $z$ position is  shown in figure~\ref{fig:Align2009_2010}.
\subsubsection{Space alignment using tracks}
\label{sec:align}
\begin{figure}[ht]
  \centering
	\subfigure[]{
	\includegraphics[width=0.7\textwidth]{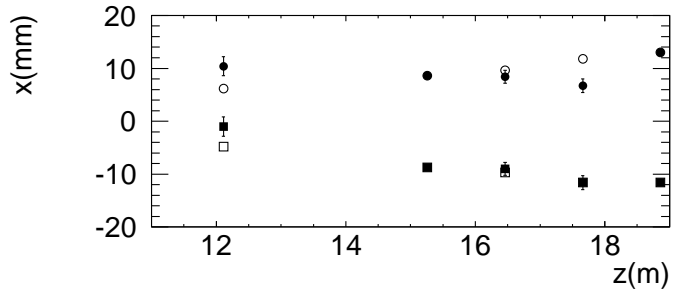}
	}
\\
	\subfigure[]{
	\includegraphics[width=0.7\textwidth]{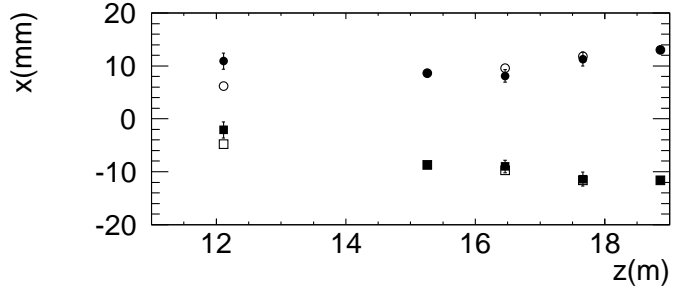}
	}
	\caption{Average position along the $x$ coordinate of the muon half stations before (a) and after  (b) the displacement   of station M4A (as described in section~\protect\ref{sec:align}). The open circles (side A) and squares (side C) represent the survey measurements while the solid circles and  squares are the positions obtained by the software local alignment with respect to the reference given by M2 and M5 stations. The errors are statistical only.}
	\label{fig:Align2009_2010}
\end{figure}

An independent determination of the position of the muon detector elements can also be obtained analyzing the tracks reconstructed in the detector. 
This is extremely useful both to check the mechanical positioning of the chambers in the stations and to monitor the alignment of the muon stations after each opening and closure.  
By studying the residual distributions between the hit and the track coordinates over the different stations it is in principle possible to determine   the detector misalignement and possibly mechanically correct it.\\
In  case the track is defined  only by the information of the muon detector, it is possible to study the relative alignment of the muon stations with respect to an arbitrary reference defined, for example, by fixing the position of two stations (\emph{local alignment}). 
As a consequence, any additional degree of freedom of the muon system like global rotation, translations or shearing can only be determined aligning the muon detector by using the tracks reconstructed also by the tracking detectors of the experiment (\emph{global alignment}).\\
The analysis described here is focused on the study of the relative positions of the muon half stations since  the statistics did not allow a precise study of single chamber alignment.
To simplify the study further, only the most relevant degrees of freedom were considered, i.e. translations in $x$ and $y$ direction.\\
The local alignment of the muon half stations was studied with respect to the half stations of M2 and M5, that were used to define the reference. Only the tracks  crossing the same side (C or A-side) of the stations were considered. 
The study was performed using two methods. In the \emph{histogram} method the tracks were defined by the straight line joining the hits found in the two reference stations. The residuals on the remaining half stations were then calculated between the clusters center and the track fit position  ($r = x_{cluster}-x_{fit}$). The mean values represent the best estimates of the alignment parameters. With the \emph{Kalman fit} method, instead, the alignment parameters were calculated iteratively by minimizing the total $\chi^2$ of the track sample with respect to the alignment parameters until convergence is reached. 
While the first method is rather simple to analyze, the second one provides a more accurate track fit accounting for multiple scattering effects. The results of the two methods were found in agreement and the Kalman fit method was eventually used for the results in the following. \\
The systematic uncertainties, mainly due to the non uniform geometrical acceptance of the detector to cosmic ray tracks and to the rather poor granularity of the detector, were estimated with Monte Carlo data in the configuration of the aligned detector. 
They amount to about 1~mm along $x$ and about 2~mm along $y$ directions.\\
The study performed on a first  data sample showed a significant displacement of the M4, A side, station along the $x$ direction of $\sim$ 5~mm, far above the systematic errors of the method, as shown in figure~\ref{fig:Align2009_2010}(a).
For this reason half station A of M4 was moved by $\sim$4~mm with respect to the other half stations M2A,M3A and M5A inside the common support to compensate the observed deviation. 
The analysis of data acquired after the displacement clearly shows the effectiveness of the correction, as one can see from figure~\ref{fig:Align2009_2010}(b).
The results showed also misalignments along the $x$ coordinate for M1 station, albeit with larger uncertainties. Since M1 halves can be easily moved independently of the other stations, it was decided to wait for more significant results with collision tracks before making adjustments. \\
The global alignment was  performed relative to  the OT detector. 
OT tracks were selected with a special setting of the track finding designed for cosmic rays and ignoring drift time information. 
Muon tracks matching the OT tracks  were selected   and  analyzed with the  Kalman fit iterative method to determine the alignment constants of the muon half stations. The reference was defined assuming the first layer of the first OT station  and the muon station M3  in their nominal position. The results are summarized in  table~\ref{tab:kalmanOTMuon}.
\begin{table}[htb] 
\caption{Misalignments $\Delta$x and $\Delta$y of muon half stations M1, M3, M4 and M5 with respect to the survey measurements calculated with the Kalman fit iterative method assuming the first layer of the first OT station and the muon station M2  in their nominal positions. The quoted errors are  statistical only. The systematic errors amount to 1 and 2~mm along the $x$ and $y$ directions.}
\label{tab:kalmanOTMuon} 
\begin{center} 
\begin{tabular}{|c| c c | c c |}
 \hline
&\multicolumn{2}{c|}{C-side}&\multicolumn{2}{c|}{A-side}\\
& $\Delta$x (mm) & $\Delta$y (mm) & $\Delta$x (mm) & $\Delta$y (mm)   \\
\hline
M1 & 1.8$\pm$0.2&--0.7$\pm$0.5&2.7$\pm$0.2&--1.8$\pm$0.5\\
M2 & -- &  -- & -- & --\\
M3 & 1.1$\pm$0.2&--1.8$\pm$0.9&--0.7$\pm$0.2&--1.3$\pm$0.9\\
M4 & 1.3$\pm$0.7&--2.8$\pm$1.1&--1.6$\pm$0.7&--2.0$\pm$1.1\\
M5 & 1.1$\pm$0.8&--1.0$\pm$1.3&2.9$\pm$0.8&--2.9$\pm$1.3\\
\hline
\end{tabular}
\end{center}
\end{table}
 The results show an overall compatibility with the survey measurements. With tracks coming from the interaction point a more accurate determination of the alignment will be possible. 


\subsection{Time alignment}
\label{subsec:timeAlig}
The muon trigger requires that particles are detected and assigned
to the proper LHC bunch crossing in each of the five muon stations. The
purpose of the time alignment is to adjust the delays of all detector
channels in order to maximize the probability that the signals fall
within the 25 ns gate around the correct bunch crossing. \\
The time alignment is achieved in a three step  procedure, the first two of which
do not rely on beam particles and are described in this paper. In the first step a pulser system sends test signals
directly to the front-end input allowing to equalize the timing of the
readout chain. In the second step, the cosmic ray tracks are used  for a
refinement of the detector internal alignment  using physical signals. Its
accuracy is limited, notably for the inner regions, by 
the available statistics and by the asynchronicity of the signals with
the LHC clock. For this reason, the ultimate alignment will be
achieved using particles from collision data, relying on the sharp arrival
time of the beam bunches.

\subsubsection{Pulser time alignment of the readout chain}
The first step of the procedure was developed to time--align all readout channels 
making use of the Pulse Distribution Module
(PDM)~\cite{bib:PDM} and of the integrated timing facilities of the
front--end (DIALOG chip) and off--detector (SYNC chip) electronics. 
The PDM received the LHC master clock 
and  generated pulse signals corresponding to
a pre-defined bunch crossing. This
pulse was distributed to the front-end electronics through the
Service Boards (SB)~\cite{bib:bocci}. From the SB the pulse was injected into the
front-end inputs and the related outputs following the normal path up to
the ODE boards, where the signal time was measured by the SYNC TDC. 
In order to make optimal use of the delay ranges available at the
DIALOG (52 ns in steps of 1.56 ns) and SYNC (175 ns in steps of 25 ns)
level, the equalization proceeded in two steps. 
In the first step (fine time alignment), the
relative time of the pulse signal with respect to the 25~ns gate was
measured by the SYNC TDC and the appropriate delays were calculated and
loaded in the DIALOGs in order to center the time spectrum 
on the 25~ns LHC gate. In the second step (coarse time alignment),
delays were applied to both the DIALOG and SYNC so that all the timing
pulses were recorded in the same pre-defined 25~ns LHC gate.
\\
The differences in the lengths of the cables bearing the pulse signals
were compensated by appropriate corrections to the delays. Cable
length values were precisely measured  during the detector cabling phase and
stored in a database. The uncertainty on these corrections limits the accuracy
of the method to a few ns. Different chamber responses and particle time of flight also
introduce misalignments that cannot be corrected by this procedure.
Moreover, due to the complexity
of the system and to the staged installation and commissioning of the
detectors, in particular of the M1 station, the procedure was
not fully achieved for the whole system before the cosmic ray data taking. \\

\subsubsection{Time alignment using cosmic rays}
\label{subsubsec:timeAligCosmic}
In this step of the procedure, cosmic ray tracks were used to align the muon detector internally 
by comparing the measurements on the same track by different detector channels.
With the reasonable assumption that all channels in a given FE board have the
same timing, the misalignment of every FE board $k$, with
respect to the other channels, was evaluated  by averaging the residual
\begin{equation}\label{eq:residual}
T_k =    t_k - \frac{\sum_{i=1}^n t_i}{n}
\end{equation}
over all tracks with a hit on FE board $k$. For each of the $n$ measurements
on a given track, the time $t_i$ is obtained
from the measured raw time $t_{\mathrm{R}}$ after correcting for the track non--projectivity as $
t =   t_{\mathrm{R}} + t_{\mathrm{PV}} - t_{\mu} $,
where $t_{\mathrm{PV}}$ is the time of
flight from the primary vertex  and $t_{\mu}$ is the actual time of
flight of the cosmic muon from an arbitrary  $z$ reference value, computed 
according to the fitted track trajectory.  The track direction is given
unambiguously by the sign of the slope in the vertical plane, as shown in figure~\ref{fig::slopes}. \\
As described in section~\ref{sec:intro} there are three different
readout configurations. In the regions where there is a
one to one correspondence between the readout channel and the logical
pad, the measurement is unambiguous. 
 For the regions with double readout (M2/3 R1/2) the $x$ and $y$ time measurements  were considered
as independent and   the time resolution was estimated separately for the two
views. For the regions where the logical pads are obtained by crossing
two $x$ and $y$ logical channels
triggered by the same physical channel, the two measurements were averaged.
In order to suppress the combinatorial background from the crossing
of two unrelated logical channel hits,  the two measurements were required 
to agree within 2 TDC bins (3.1 ns) for regions with single readout,
and within 25 TDC bins (39 ns, more than 9 times the expected
resolution) for regions with double readout. \\
Signals recorded in the first two and
last two bins of the 16 bin range of the TDC were not used in this analysis,
in order to remove the effect of an unwanted non--linear behavior of the
TDC, which distorts  the TDC spectra for signals
falling near the borders of the LHC time gate. Even with this
cut, some residual effect was left and was taken care of in the
measurement of the time resolution, as described in
section~\ref{subsec:timeRes}.\\ 
For each FE board a correction was computed as $
\Delta T_k = - \alpha  <T_k> $ and the procedure was iterated ($\alpha=0.8$ is a factor to damp
possible oscillations) until the fraction of statistically significant corrections
became negligible ($<$1\%). In case of hit clusters, all of them were used  for the first three
iterations, and only the first in time  for the next ones, in order 
  to suppress the effect of delayed cross--talk.
The procedure converged after six iterations. \\
This calibration was limited by statistics for the inner regions. 
Though the average number of tracks recorded per FE board  is over 50, the
value is much smaller for the inner
regions, notably for M1R1 (3.5 tracks/FE board on average). In order to avoid smearing 
the timing calibration with large statistical fluctuations,
we applied the average correction of the corresponding region to 
the  channels for which  the computed correction was not significant.\\
The alignment with cosmic ray tracks allowed to identify and fix several problems of the pulse distribution
cable chains. Systematic delays among stations of up to 10 ns 
were corrected, and smaller biases among sides or ODEs inside the same
station were identified. The r.m.s of the statistically significant corrections
amounts to 6.5 ns. \\
A second  time alignment procedure using cosmic rays was developed independently,
consisting in measuring the average channel delays with respect to
the calorimeter trigger.
In this alternative analysis, the channels, instead of being grouped
by FE board, were divided in spatial regions containing chambers with the
same characteristics. Those regions are large enough to provide
statistically significant track samples but are not guaranteed to contain
all channels with the same timing. 
Nevertheless the comparison of the two sets of constants shows an  excellent agreement (85\% correlation factor). 
The second procedure was used to
compute the overall time shift needed to align the muon detector
with the L0 trigger, while the previous stand--alone procedure was preferred for the
relative alignment, in order not to be biased by possible
imperfections of the calorimeter internal alignment, that was also
being refined. \\
The stability of the alignment corrections was also checked by repeating the 
calibration with different cuts against the cross--talk and  TDC non--linearity effects. 
No variations larger than 2 ns were observed.
\section{Detector performance}
\subsection{Total efficiency}
\label{sec:effi}

\begin{figure}[ht]
\centering{
\includegraphics[width=0.6\textwidth,height=0.6\textwidth]{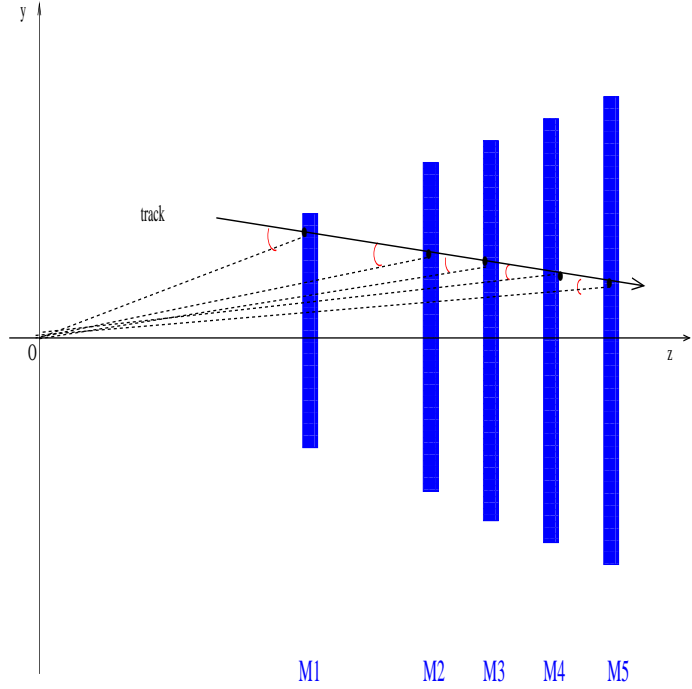}
\caption{
View (not to scale) of the LHCb muon detector 
on the $yz$ plane. The projected angles $\theta^p_{i,yz}$ 
for the five stations are shown.
The drawing is not to scale.
}\label{fig::angle_drawing}			 
}
\end{figure}

The efficiency calculation was performed  with a re--run of the neural network based pattern recognition and track reconstruction, removing all hits of one station at the time. Then the number of tracks for which a hit was  found inside a window of 6$\times$4 logical pads from the extrapolated point in that station divided by the total number of track candidates provided the efficiency values. 
The  efficiencies  obtained in this way were of the order of 85\% on average, with the  major source of inefficiency being the non hermeticity of the detector for tracks not coming from the interaction point.
In order to get rid of this effect the efficiency was studied as a 
function of the angle $\theta^p_{i,x(y)z}$ ($i=1,...5$)
defined for each station M$_{i}$ as the 
projected angle the track forms with 
the line connecting the hit with the interaction 
point (as illustrated in figure~\ref{fig::angle_drawing}
for the $yz$ projection). 
\begin{figure}[ht]

  \centering
	\subfigure[]{
	\includegraphics[width=1.\textwidth]{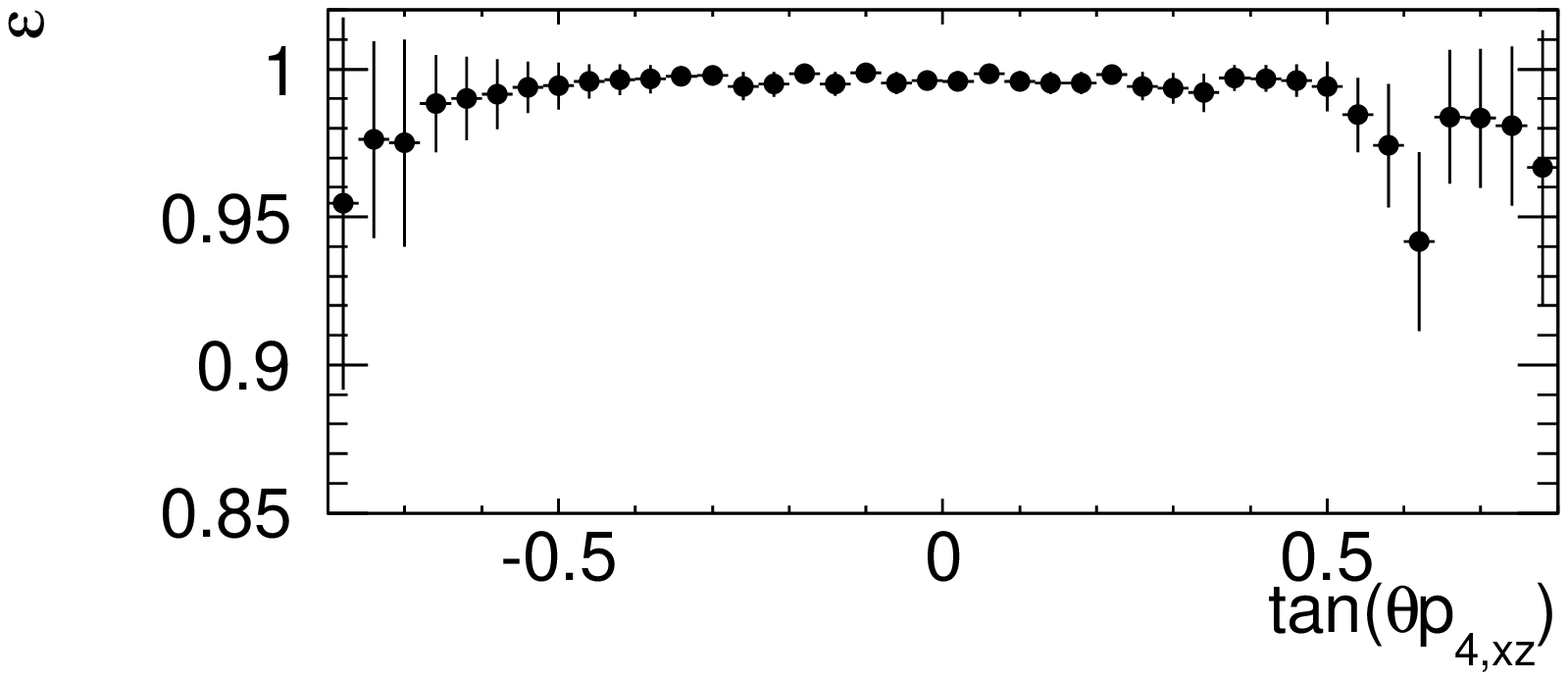}
	}\\
        \subfigure[]{
	\includegraphics[width=1.\textwidth]{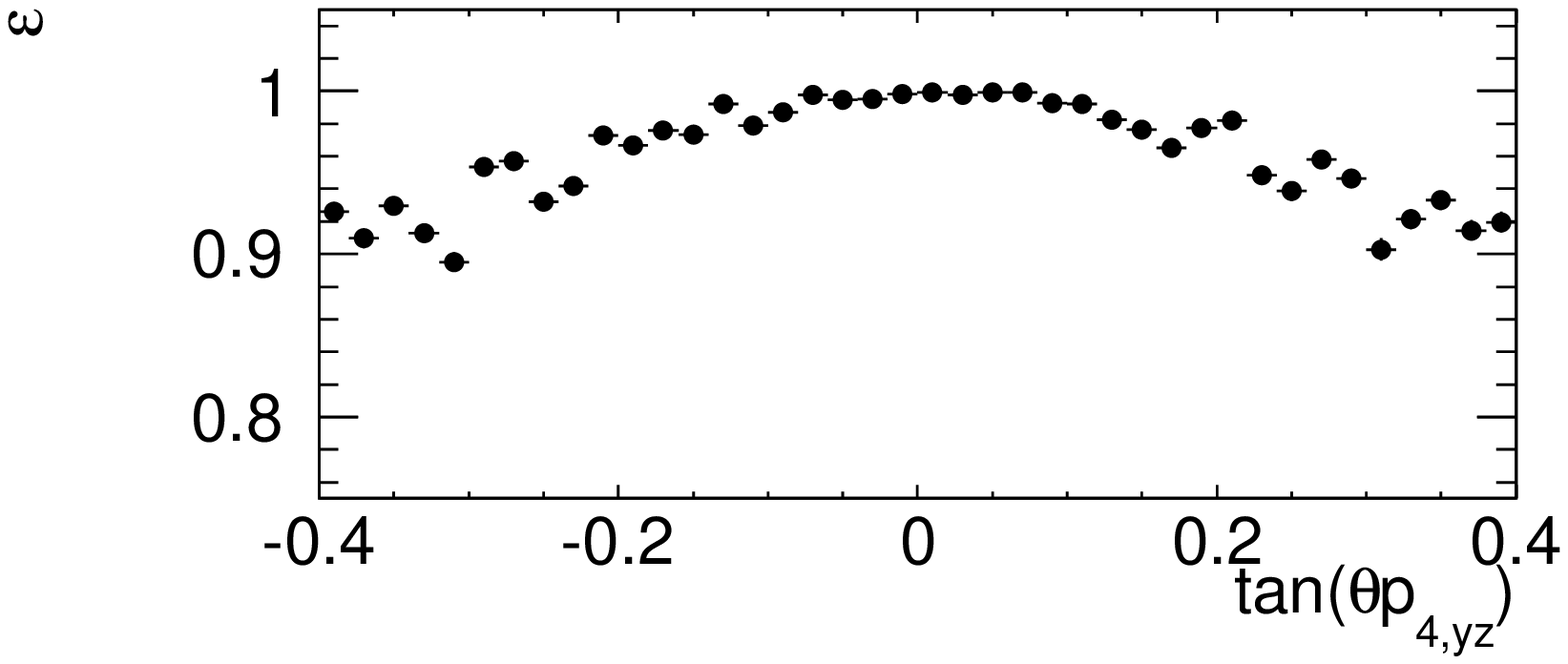}
	} 

 \caption{Efficiency $\epsilon$ as a function of  $\theta^p_{4,xz}$ (with $tan(\theta^p_{4,yz})< 0.06$) (a) and $\theta^p_{4,yz}$ (with $tan(\theta^p_{4,xz})< 0.2$)(b) angle for station M4, as defined in figure~\protect\ref{fig::angle_drawing}.} 
 \label{fig:effivsangle}

\end{figure}
The efficiency was calculated as a function 
of $\theta^p_{i,x(y)z}$ for the all  stations, averaging the measurements of the four regions, to increase the statistical significance of the measurement.  
Since cosmic rays can be absorbed before reaching the outer stations M1 and M5 simulating   chamber inefficiency,  for the efficiency measurement of M1 and M5 only forward and backward tracks were used, respectively, and a fiducial region around the outer and inner regions of the detector was defined.
The results for station M4 are shown in figure~\ref{fig:effivsangle}.  The efficiency reaches a plateau value at  small values of tan$(\theta^p_{i,yz})$ for all stations while a  flatter dependence on  tan$(\theta^p_{i,xz})$ is observed, explained  by the smaller $y$ size than $x$ size of chambers, leading to a smaller hermeticity in the $yz$ plane.  Table~\ref{tab::efffit2} quotes the measured efficiencies for the five stations  integrated over tan$(\theta^p_{i,yz})< 0.06$  and  tan$(\theta^p_{i,xz})< 0.2$.  The latter cut was actually  applied to define a fiducial region around zero angle also in the $xz$ plane but, due to the quite flat dependence of the efficiency on  tan$(\theta^p_{i,xz})$ , has some degree of arbitrariness. Still the measured efficiencies are stable within  0.1~\% with respect to the precise cut values.
\begin{table}[ht]
\caption{
Total efficiency $\epsilon$(\%)  (in the wide gate) for tracks having tan$(\theta^p_{i,xz})< 0.2$ and  tan$(\theta^p_{i,yz})< 0.06$ 
for the five stations of the LHCb muon detector.}
\label{tab::efffit2}
\begin{center}
\begin{tabular}{|c|c|c|c|c|c|} \hline
 & M1 & M2  & M3   & M4  & M5  \\ \hline
 $\epsilon$(\%) &   98.8$\pm$0.4
                                 &   99.7$\pm$0.1
                                  &   99.9$\pm$0.1
                                  &   99.8$\pm$0.1
                                  &   99.8$\pm$0.1 \\
  \hline
\end{tabular}
\end{center}
\end{table}

\subsection{Cluster size}
\label{sec:AveragePadMult}
\begin{figure}[ht]

  \centering
	\subfigure[]{
	\includegraphics[width=0.46\textwidth]{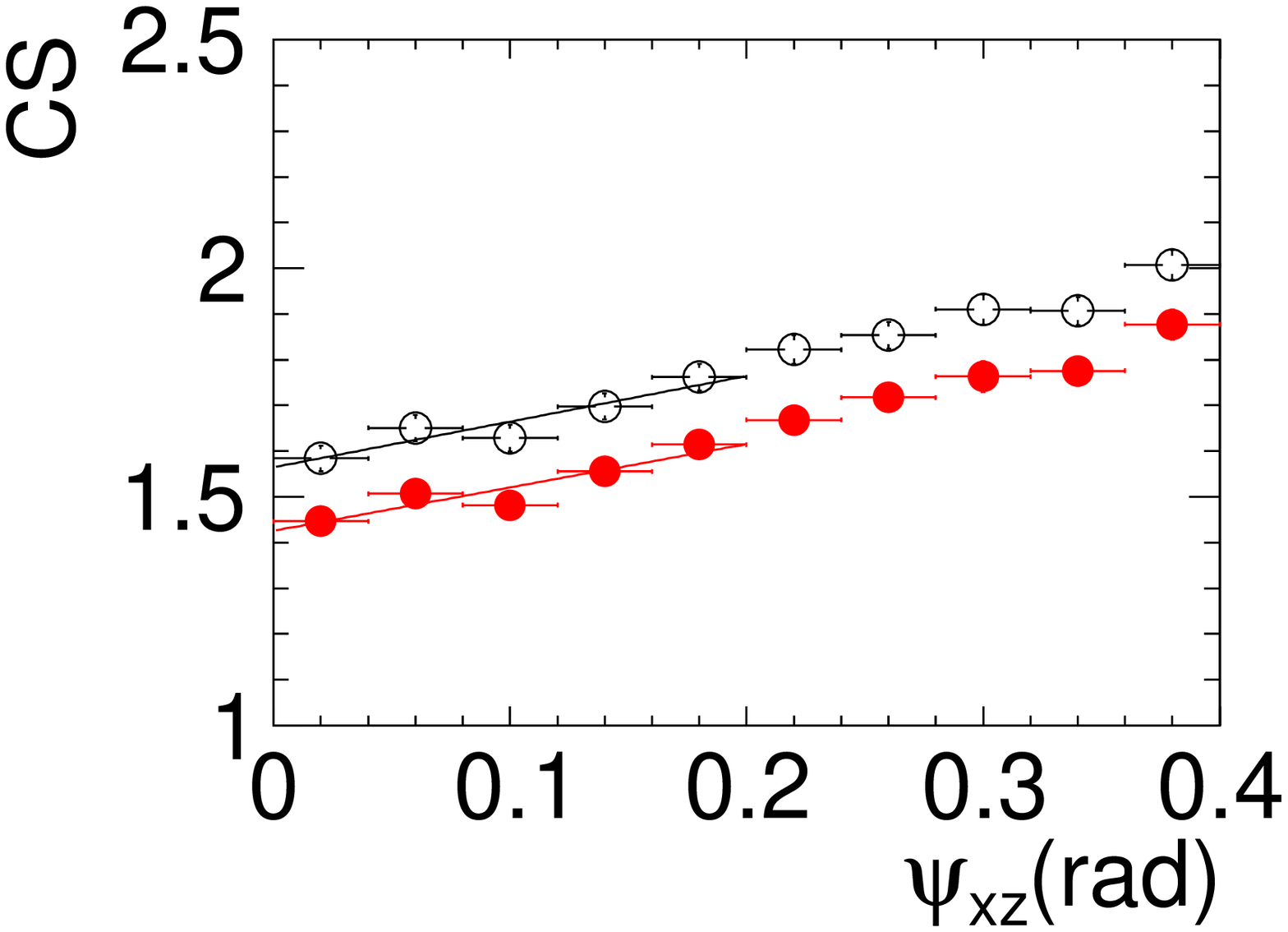}
	}
        \subfigure[]{
	\includegraphics[width=0.46\textwidth]{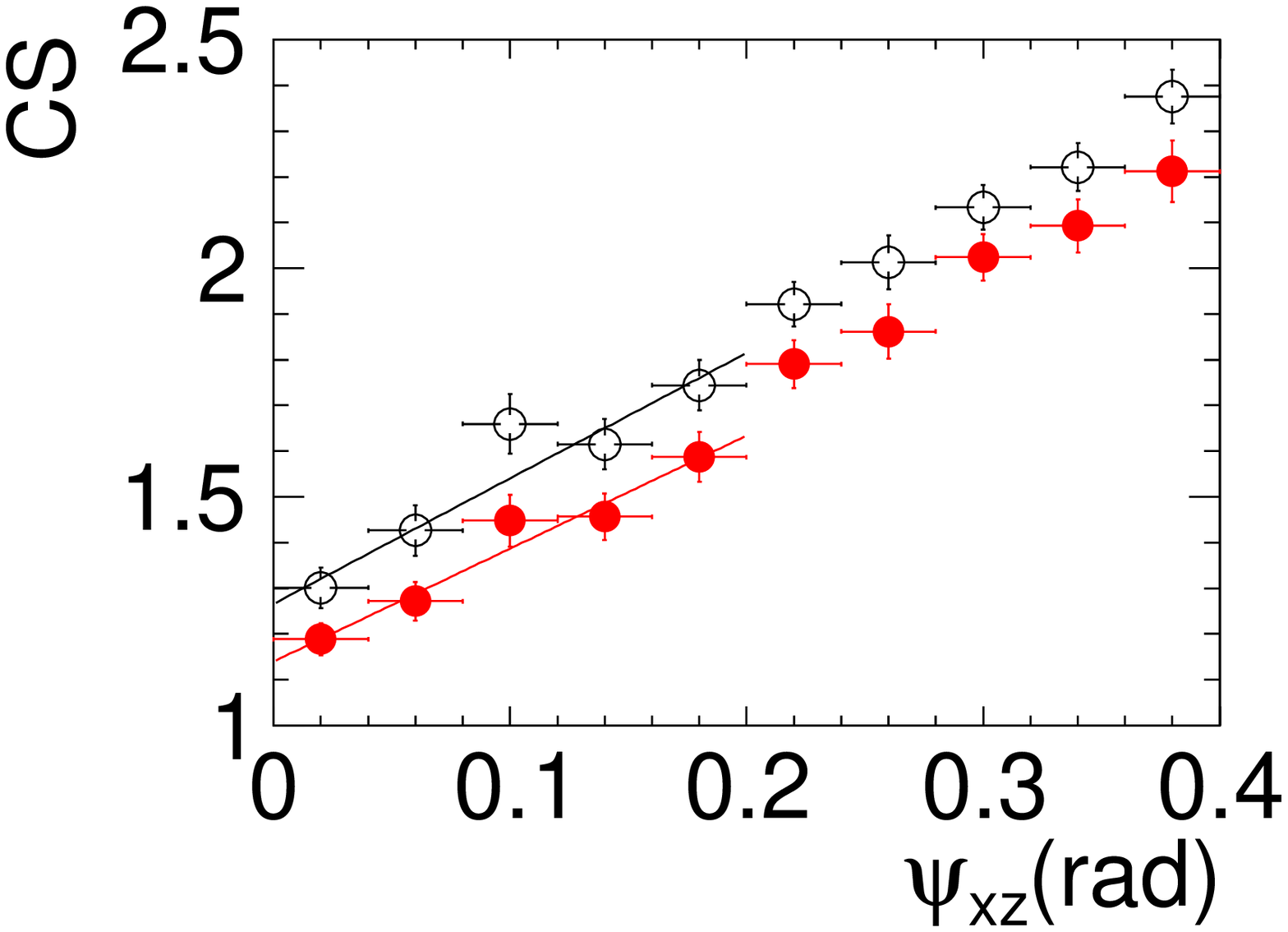}
	} \\
	\subfigure[]{
	\includegraphics[width=0.46\textwidth]{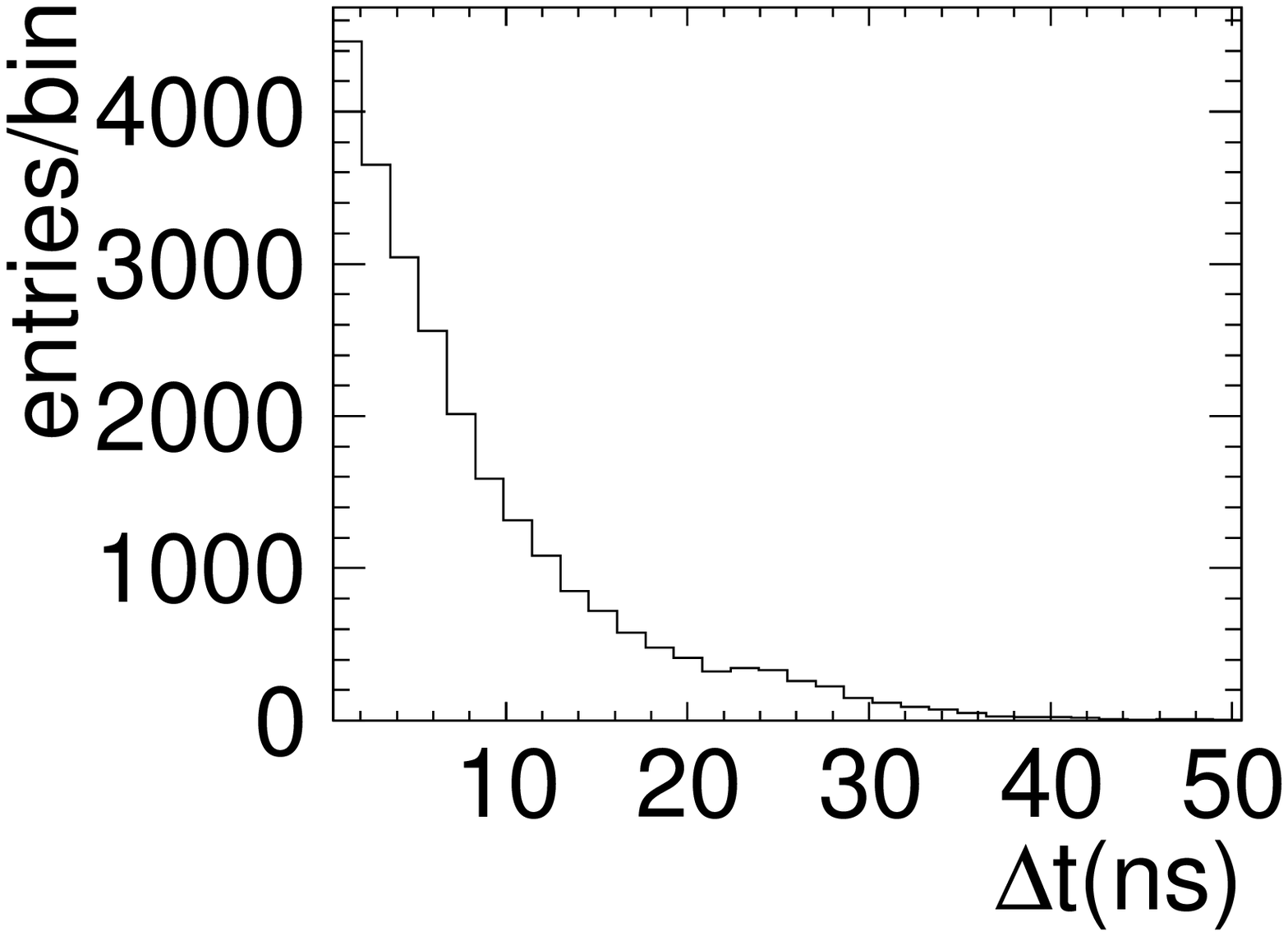}
	} 
        \subfigure[]{
	\includegraphics[width=0.46\textwidth]{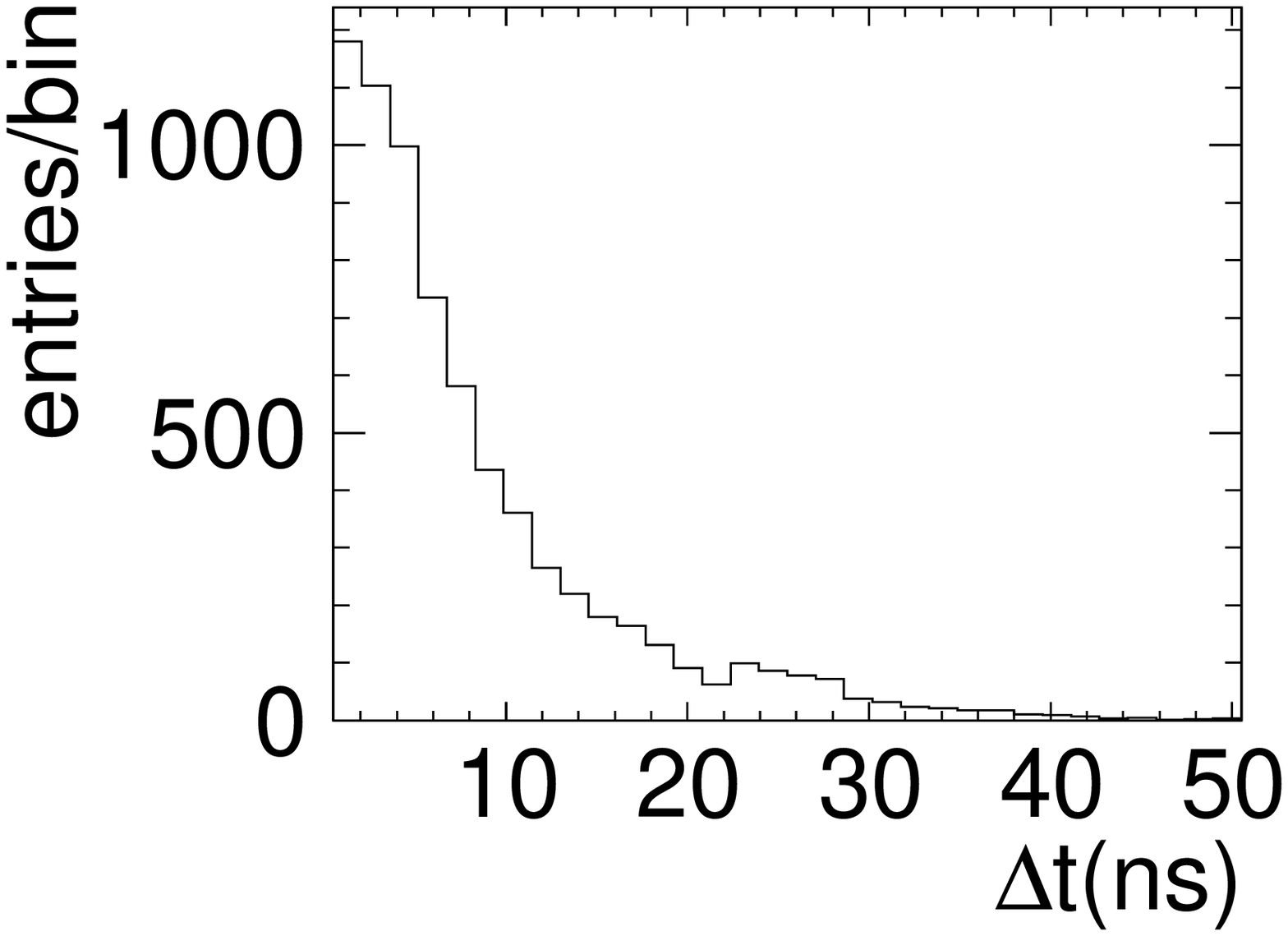}
	} 

 \caption{{ Average pad cluster size  vs. $\psi_{xz}$} for $|\psi_{yz}|<0.5$~rad in the wide gate (open circles) and in the  25~ns LHC gate (solid circles), for M2R3 (a) and M3R2 (b) chambers. Time ordered difference between the  $x$ logical channels forming the cluster  and the one occurring first in time, in events with cluster size  larger than one, for M2R3 (c) and M3R2 (d) chambers. } 
 \label{fig:ybins}

\end{figure}

The track finding procedure of the muon trigger algorithm is based on logical pad signals which are combined one per station to form five hit tracks, with the logical pads having a smaller size in $x$ to measure the transverse momentum, as  previously explained. To avoid spoiling the transverse momentum measurement and to limit the number of hit combinations (the trigger algorithm does not consider clusters) it is therefore important to minimize the  cluster size (CS) along $x$, i.e. the number of adjacent logical pads along $x$ fired by the same track.
 The cluster size is an intrinsic characteristic of a chamber, but it is also affected by track inclination, given the non negligible thickness of the multi--gap chambers. To distinguish between the two effects, the cluster size was measured as a function of the angle $\psi_{xz}$ that the muon track makes with the perpendicular to the chamber. \\ 
Figure~\ref{fig:ybins}  shows the  cluster size   vs. $\psi_{xz}$  for tracks with $|\psi_{yz}|<0.5$~rad for M2R3 (a) and M3R2 (b) chambers.  In regions where the logical pads are smaller, a steeper dependence on $\psi_{xz}$ is observed, as expected from the geometry. 

\begin{figure}[ht]
  \centering
	\subfigure[]{
	\includegraphics[width=0.5\textwidth]{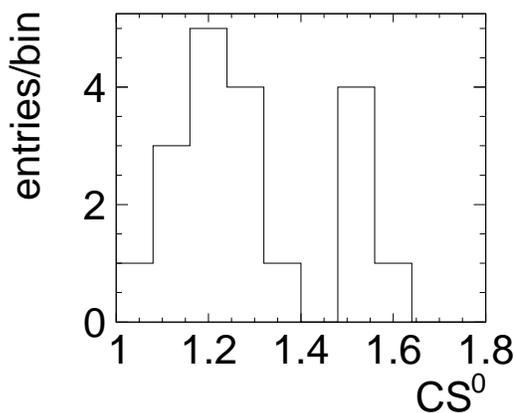}
	}\\
        \subfigure[]{
	\includegraphics[width=0.5\textwidth]{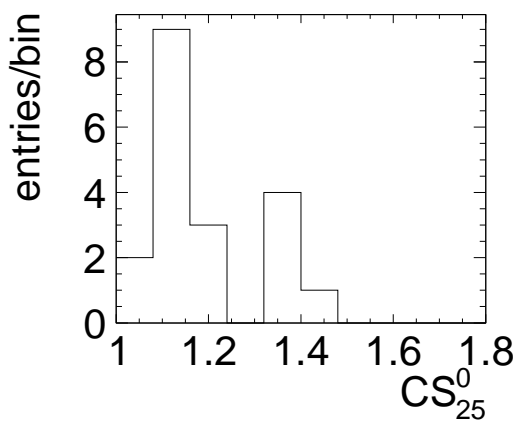}
	} 
    \caption{Logical pad cluster size at $\psi_{xz}\to$0~rad, $CS^0$, for  tracks with $\psi_{yz}<$0.5~rad  in an infinite time window $CS$ (a)  and in the 25~ns LHC gate $CS^0_{25}$ (b). One entry per chamber type. }  
    \label{fig:clsize}
\end{figure}

The hits affecting the muon system performance in collision mode are only those occurring during the 25~ns LHC gate. Therefore our time integrated cluster size measurement is an over estimate of the effect. The cluster size in the 25~ns LHC gate, $CS_{25}$ was  derived by convoluting the measured time distribution of the main hit, as described in section~\ref{subsec:timeRes}, centred in the  25~ns LHC gate, 
with  the experimental distribution of the time ordered difference between the  $x$ logical channels forming the cluster  and the one occurring first in time, in events with cluster size  larger than one, shown in figure~\ref{fig:ybins} (c) for M2R3 and (d) for  M3R2 chambers. 
Figure~\ref{fig:ybins} shows the  corrected cluster size vs. $\psi_{xz}$   for $|\psi_{yz}|<0.5$~rad for M2R3 (a) and M3R2 (b) chambers. 
The first five  open  points of figure~\ref{fig:ybins}  were fitted with  a straight line and  the extrapolation of this line to zero angle, $CS^0$,  was reported in figure~\ref{fig:clsize}(a), with one entry per chamber type. The typical uncertainty  on $CS^0$  is of the order 0.05 and is due to the uncertainty from the extrapolation.
The first five   solid  points of the same figure  were fitted with  a straight line and  the extrapolation of this line to zero angle, $CS^0_{25}$,  was reported   in figure~\ref{fig:clsize}(b), with one entry per chamber type.
Most chamber types are well inside the specifications; for those stations and regions  where  $CS^0_{25}$  is at the edges,  there is a plan for the future running of the experiment to lower  the high voltage value, which will bring the cluster size to even smaller values. 

\subsection{Time resolution}
\label{subsec:timeRes}

\begin{figure}[ht]
  \centering
	\subfigure[]{
	\includegraphics[width=0.7\textwidth]{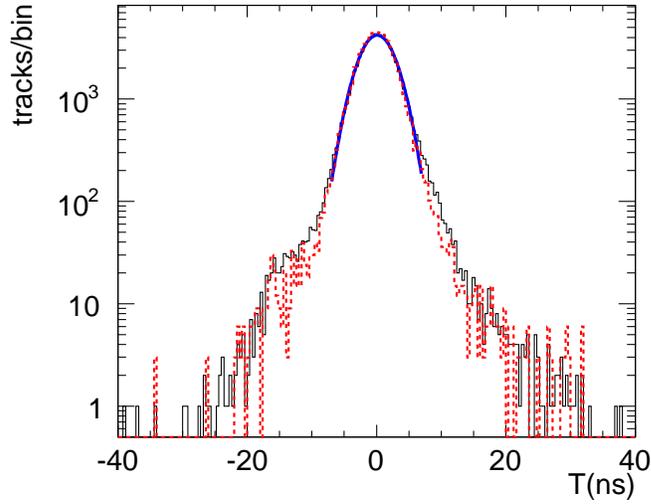}
	}\\
        \subfigure[]{
	\includegraphics[width=0.7\textwidth]{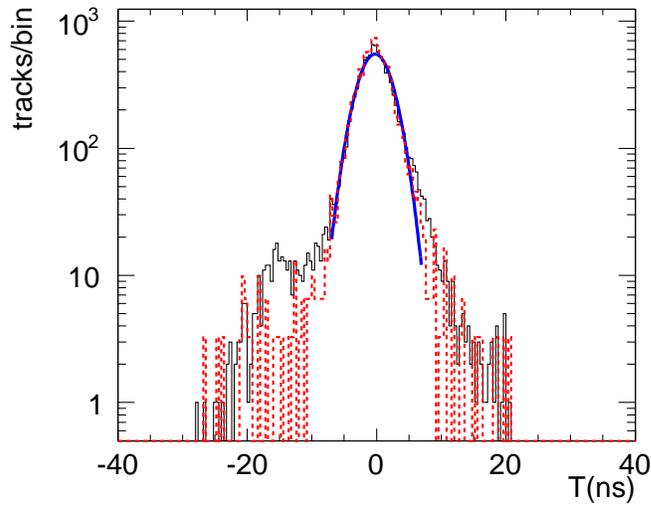}
	} 
    \caption{{ Distribution of time residuals $T$ for M5R4 (a) and M3R2 $x$ view
        (b) chambers. For M5R4 chambers the same physical signals is
        readout by two TDC channels for the $x$ and $y$ views, that were
        required to agree. For the M3R2 chambers there are two
        independent readouts for the two views.
        The dotted distribution is obtained after selecting
        tracks centered in the 25~ns LHC gate and is normalized to the same area
        of the full plot to show the effect on the tails.}} 
    \label{fig:timeResid}
\end{figure}

The detector time resolution was estimated for each region from the distributions of the
time residual~\ref{eq:residual} after applying the corrections for
time misalignments described in section~\ref{subsec:timeAlig}. We used
a  data sample different, though acquired in similar conditions, from the one used
for the time alignment, in order to avoid a bias on the time
resolution from the over--training of the alignment. \\
The residual distributions exhibit some non
gaussian tails.  A detailed study was performed comparing regions characterized by a different readout method. An example is shown in figure~\ref{fig:timeResid}. Tails are 
smaller for regions having the same physical
signal read by two logical channels (for $x$ and $y$ views), where the
coherence of the two measurements was required, as in
figure~\ref{fig:timeResid}(a). This
suggests that tails are due to the TDC misbehavior at the  25~ns LHC gate borders and not to the
intrinsic chamber response. As a further test of this hypothesis, the residuals were looked at, 
after selecting only forward tracks whose absolute time
extrapolated at calorimeter, measured by the hits other than the one
under scrutiny, is within $\pm$ 6 ns from the LHC gate center.
For those tracks, signals  are expected
to be more centered on the gate. The effect of this cut is shown on
figure~\ref{fig:timeResid} for regions M5R4 and M3R2x. 
Since non--gaussian tails are mostly an artifact of the readout,
the resolution was eventually  estimated from a gaussian fit performed on the core of the
distribution of time residuals in region $R$ for tracks with $n=5$
measurements:
\begin{equation}\label{eq:sigmacore}
s_{Rcore} = \frac{n}{n-1}  \sqrt{
 \sigma_R^2(\mathrm{resid})-\frac{n-1}{n^2} \overline{\sigma}^2}
\end{equation}
where $\sigma_R(\mathrm{resid})$ is the fit result and
$\overline{\sigma}$ is the average resolution estimated in the 
same way: \\
$ \overline{\sigma}~=~\sqrt{n/(n-1)}~~~\sigma (\mathrm{all~residuals})$. \\
\begin{figure}[ht]
      \centering
      \includegraphics[width=0.7\textwidth]{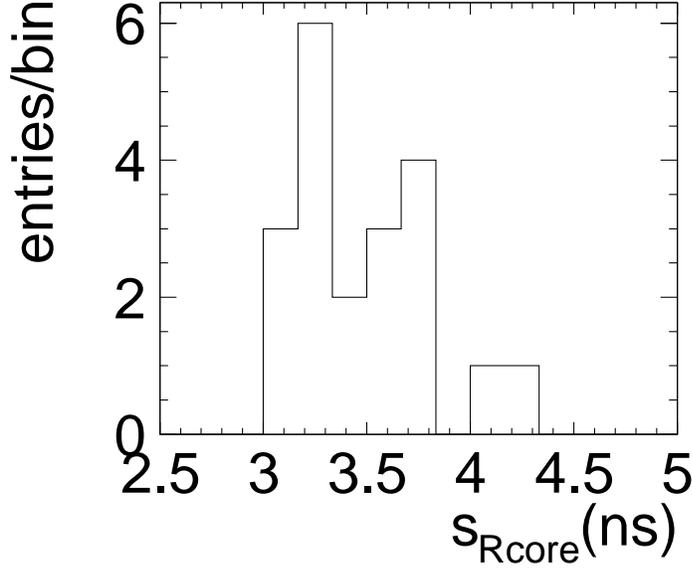}
\caption{Resolution measurements $s_{Rcore}$ , in ns, after correcting for the systematic effects. One entry corresponds to one chamber type; the chambers with double readout have two entries.}
      \label{fig:timeRes}
\end{figure}
\begin{table}[ht]
  \caption{ Resolution measurements, in ns, before  ($s_{Rcore}$) and
   after correcting for the systematic effects for the different chamber types.} 
\label{tab:timeRes}
\begin{center}
  \begin{tabular}{|r|c|c|c|c|}
    \hline
chamber type &  $s_{Rcore}$          &  final \\ 
       &                      &  estimate\\
    \hline
    M1R1  &  5.0  $\pm$  0.5 &   4.3 $\pm$ 1.2\\
    M1R2  &  4.0  $\pm$  0.4 &  3.4 $\pm$ 0.6\\
    M1R3  &  3.9  $\pm$  0.1 & 3.7 $\pm$ 0.2\\
    M1R4  &  4.0  $\pm$  0.1 &  3.8 $\pm$ 0.2\\
   M2R1x  &  3.0  $\pm$  0.2 &   3.1 $\pm$ 0.3\\
   M2R1y  &  3.0  $\pm$  0.2 &   3.1 $\pm$ 0.3\\
   M2R2x  &  3.0  $\pm$  0.1 &  3.2 $\pm$ 0.2\\
   M2R2y  &  3.0  $\pm$  0.1 &   3.2 $\pm$ 0.2\\
    M2R3  &  3.3  $\pm$  0.1 &  3.5 $\pm$ 0.2\\
    M2R4  &  3.3  $\pm$  0.1 &   3.3 $\pm$ 0.1\\
   M3R1x  &  3.4  $\pm$  0.2 &   3.6 $\pm$ 0.3\\
   M3R1y  &  3.2  $\pm$  0.2 &   3.4 $\pm$ 0.3\\
   M3R2x  &  3.0  $\pm$  0.1 &   3.2 $\pm$ 0.2\\
   M3R2y  &  3.0  $\pm$  0.1 &  3.2 $\pm$ 0.2\\
    M3R3  &  3.3  $\pm$  0.1 &   3.5 $\pm$ 0.2\\
    M3R4  &  3.1  $\pm$  0.1 &    3.1 $\pm$ 0.1\\
    M4R1  &  4.1  $\pm$  0.3 &    4.1 $\pm$ 0.4\\
    M4R2  &  3.6  $\pm$  0.1 &   3.7 $\pm$ 0.2\\
    M4R3  &  3.7  $\pm$  0.1 &    3.7 $\pm$ 0.1\\
    M4R4  &  3.2  $\pm$  0.1 &   3.3 $\pm$ 0.2\\
    M5R1  &  3.3  $\pm$  0.3 &   3.3 $\pm$ 0.3\\
    M5R2  &  3.4  $\pm$  0.2 &   3.5 $\pm$ 0.3\\
    M5R3  &  3.5  $\pm$  0.1 &    3.5 $\pm$ 0.1\\
    M5R4  &  3.3  $\pm$  0.1 &     3.3 $\pm$ 0.1\\
\hline

\end{tabular}
\end{center}
\end{table}
Several systematic effects affecting the resolution measurement have
been studied with the help of a toy Monte Carlo, such as the  removal of  signals at the 2+2 border TDC bins,
the selection of the first cluster hit in time, the residual time misalignments due to lack of statistics in some regions and the fit procedure. \\  
The total systematic
corrections turned out to be of the order of  0.1-0.2~ns and only for M1R1 and M2R2 amount to -0.7 and -0.6~ns, respectively. Figure~\ref{fig:timeRes} , with one entry per chamber type, 
shows the final estimated resolution. For internal reference of the LHCb muon collaboration the results are also detailed in table~\ref{tab:timeRes}, where the first column shows the $s_{Rcore}$  resolution and the second column shows the final estimation after the systematic correction. The typical uncertainty  on the single  data point is of the order of 0.2~ns, including both statistical an systematic contributions.  \\
It can be noticed  that the time resolution results lie in the range between 3 and 4~ns; only two chamber types have time resolutions worse than 4~ns and it is due to  residual  time misalignment due to lack of statistics in the inner regions. \\
As discussed in section~\ref{sec:detreq} , with the cosmic data it was not possible to directly measure the 20~ns efficiency. Rather, from the  measured  values of time resolution and with the results of figure~\ref{fig:EffRMSvsThr}, an indirect estimate could be obtained, assuming the time resolution to be  the only source of inefficiency, and showed that, apart from  the triple-GEM M1R1 chambers, a 20ns efficiency  above 97.5\% was archived. \\
A direct measurement of the 20~ns efficiency is going to be discussed in a future paper describing muon system performance  with LHC beams.

\section{Conclusions}

A  study of the LHCb muon system performance was presented, using  cosmic ray data taken during the year 2009.   The space and time alignment and the measurement of chamber total efficiency, time resolution and cluster size were discussed. The results confirm the expected detector performance. 

\section{Acknowledgments}

The successful construction, installation and operation of the muon system would not have been possible without 
the technical support in each of the participating laboratories:
we wish in particular to thank G.~Paoluzzi, A.~Salamon of INFN, Roma Tor Vergata,
F.~Maletta of INFN, Firenze, L.~LaDelfa, D.~Marras and M.~Tuveri of INFN, Cagliari, W.~Rinaldi and F.~Iacoangeli of INFN, Roma. \\
We would like to thank  the LHCb data acquisition team and the LHCb Calorimeter and Outer Tracker groups. \\
We also wish to thank the team of surveyors at CERN, in particular S. Berni and J.C. Gayde, for their important contribution to the detector alignment.

\end{document}